# AFM-IR of EHD-Printed PbS Quantum Dots: Quantifying Ligand Exchange at the Nanoscale

Lorenzo J. A. Ferraresi[a,b,c], Gökhan Kara[a,b], Nancy A. Burnham[e,f], Roman Furrer[b], Dmitry N. Dirin[c,g], Fabio La Mattina[b], Maksym V. Kovalenko[c,g], Michel Calame[b,d], Ivan Shorubalko[b]

*(a) These authors contributed equally*

*(b) Laboratory for Transport at Nanoscale Interfaces, Empa – Swiss Federal Laboratories for Materials Science and Technology, CH-8600 Dübendorf, Switzerland*

*(c) Department of Chemistry and Applied Biosciences, ETH – Swiss Federal Institute of Technology Zurich, CH-8093 Zurich, Switzerland*

*(d) Department of Physics and Swiss Nanoscience Institute, University of Basel, CH-4056 Basel, Switzerland*

*(e) Departments of Physics and Biomedical Engineering, Worcester Polytechnic Institute, Worcester, Massachusetts*

*(f) Concrete and Asphalt Laboratory, Empa, Swiss Federal Laboratories for Materials Science and Technology, Dübendorf, Switzerland*

*(g) Laboratory for Thin Films and Photovoltaics, Empa – Swiss Federal Laboratories for Materials Science and Technology, CH-8600 Dübendorf, Switzerland*

**Abstract**

Colloidal quantum dots (cQDs) recently emerged as building blocks for semiconductor materials with tuneable properties. Electro-hydrodynamic printing can be used to obtain sub-micrometre patterns of cQDs without elaborate and aggressive photolithography steps. Post-deposition ligand exchange is necessary for the introduction of new functionalities into cQD solids. However, achieving a complete bulk exchange is challenging and conventional infrared spectroscopy lacks the required spatial resolution. Infrared nanospectroscopy (AFM-IR) enables quantitative analysis of the evolution of vibrational signals and structural topography on the nanometre scale upon ligand substitution on lead sulphide (PbS) cQDs. A solution of ethane-dithiol in acetonitrile demonstrated rapid (~60 s) and controllable exchange of approximately 90% of the ligands, encompassing structures up to ~800 nm in thickness. Prolonged exposures (>1 h) led to the degradation of the microstructures, with a systematic removal of cQDs regulated by surface-to-bulk ratios and solvent interactions. This study establishes a method for the development of devices through a combination of tuneable photoactive materials, additive manufacturing of microstructures, and their quantitative nanometre-scale analysis.

**Introduction**

Colloidal quantum dots can be synthesised in and deposited from solutions into photosensitive films with semiconducting behaviour. Their properties can be tailored through the choice of the core material, the core size, and the surface passivation.[1] Together with the ease of fabrication, the tuneable properties represent a great advantage for the fabrication of diverse devices, including photodetectors.[2] Both material research and device engineering are needed for colloidal quantum dots to be integrated in an increasing number of applications.

Lead sulphide cQDs have been extensively studied thanks to their size-tuneable spectral sensitivity in the near and short-wave infrared, as well as to the relatively simple synthesis routes.[3–6] As-synthesised cQDs are commonly dispersed in non-polar solvents with long, insulating ligands. These can be removed and substituted, with the choice of the final molecule taking into account surface passivation[7] and energy-level tuning[8], together with inter-particle distance[9], determining the transport properties of the material. Two main methods exist to perform this molecular substitution: liquid-phase ligand exchange (LPLE) and solid-state ligand exchange (SSLE). They are defined based on whether the exchange happens respectively before or after the deposition of the cQD film.

In the case of SSLE the penetration of ligands into the solid cQD film is limited, necessitating a layer-by-layer fabrication. Typically, a few-nanometre layer of cQDs is deposited and then ligand-exchanged, with these steps repeated to obtain a film with the required thickness.[6,10–12] This laborious fabrication shifted attention to the LPLE, where the careful optimisation of concentrations and solvents led to single-step depositions of ligand-exchanged cQD films.[13] Spin-coating, the deposition method of choice in both cases, results in large material waste, worsened by a layer-by-layer deposition.[14] Further complexities arise once the active layer is structured and integrated into devices: photolithography, widely used in fabrication processes, can easily harm the active layer by exposing it to aggressive gaseous or liquid environments.[15]

Microstructures of colloidal semiconductors can be additively manufactured through printable active inks, drastically improving the material economy. The best printing spatial resolution was obtained through electrohydrodynamic (EHD) printing, where the application of voltage pulses to the ink enables the precise fabrication of micrometre[16] and sub-micrometre[17] patterns. This requires careful optimisation of the ink properties, including its conductivity, viscosity, and vapour pressure.[18] Devices have already been fabricated at lower resolutions via inkjet-printing, following both SSLE[19] and LPLE[20] procedures. The absence of photolithographic structuring avoided damage to the active layer and also overcame limits to device design.

The higher resolution of EHD printing permits downscaling devices to few micrometres. Understanding the impact of ligand exchange processes on these printed microstructures is critical for imparting needed functionalities. The key parameters to be monitored in this process are the total volume of structures and the residual presence of native ligands. Chemical treatments can damage the network of inorganic quantum dots and organic ligands, causing losses of photoactive material and associated volume variation. If cQDs are not lost, the change in volume is proportional to changes in the ligand shell. Conventional IR spectroscopy allows us to associate the volume variations with the exchange of ligands through the intensity of their unique vibrational signatures.[6,21–23] Nevertheless, its lateral resolution is limited to micrometres by optical aberration. In contrast, the near-field AFM-IR technique provides for parallel analysis of both structural topography and chemical composition at the nanoscale. [24]

In this work, a method is developed to enable the quantitative study of ligand exchange processes at the nanoscale. Both the variations of volume and of chemical composition are considered simultaneously as measured through AFM-IR in tapping mode. Microstructures are fabricated by EHD printing of PbS cQDs with oleic acid (OA) ligands in a non-polar solvent (tetradecane) for the best results in term of spatial resolution. A one-step SSLE is then applied to replace the insulating OA with ethane-dithiol (EDT) in acetonitrile (ACN). Microstructures (area nominally 10 x 10 μm$^2$) with different heights (125±20, 470±60, 745±90 nm) are printed on each of the samples. Different samples are then exposed to the ligand solution for different time intervals (60 seconds, 1 hour, and 12 hours). An additional sample is used to evaluate the effect of overnight exposure to the ACN solvent only, decoupling it from the effect of EDT ligands. The high surface-to-volume ratio of the printed structures enhances the interaction with the ligand solution: the most effective ligand exchange is obtained in the surprisingly short time of 60 seconds, independently of thickness. With an increasing treatment time, the removal of OA is not improved, while the ligand solution damages the active layer, removing quantum dots and damaging structures through cracks.

**Results and discussion**

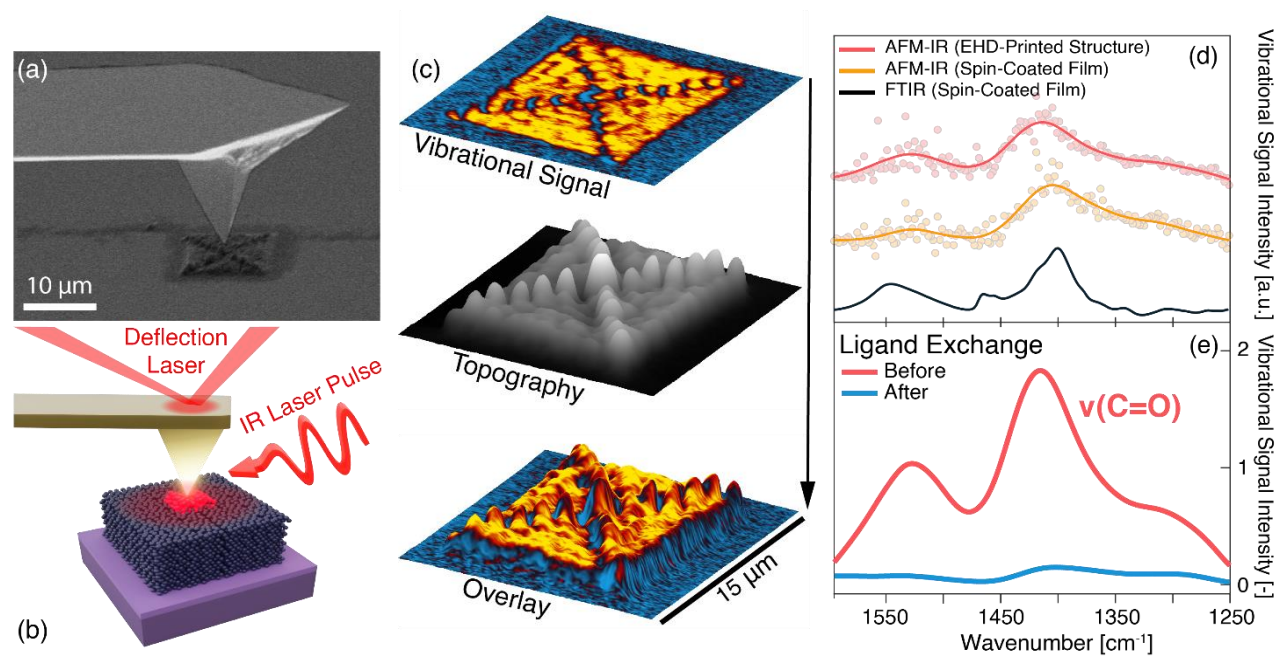

**Figure 1:** AFM-IR principle and detection of OA removal. (a) SEM images of the tip used in this AFM-IR analysis and of the sample with the printed cQD microstructures (to scale, pictures were collected separately). (b) Schematic illustration of the AFM-IR technique (not to scale). The metallic tip enhances the incoming IR laser pulses (5-11 μm or 900-1900 cm$^{-1}$). The IR light is absorbed through the vibrational modes of organic ligands, causing a pulsed thermal expansion detected mechanically by the AFM tip. The deflection laser is used for the conventional topography measurement. (c) Results from AFM-IR scans of structures, where topography and chemical composition maps are combined to clearly locate targeted molecules on the surface. (d) Spectra of the investigated PbS cQDs with oleic acid (OA) ligands. The reference vibrational signal associated with OA's carboxylic group (black) is measured through FTIR from a spin-coated film. AFM-IR measurements of EHD-printed structure (red) and spin-coated sample (orange) are shown, no difference in the AFM-IR signal can be observed based on the deposition method. (e) AFM-IR spectra collected from a 745 nm thick printed structure. The technique clearly

detects both the as-printed IR signal (red) and the quenched signal once the ligand exchange treatment is applied (blue). Both signals have been normalised to a reference sample, as described in Methods.

The ligand exchange process determines several properties of the final colloidal semiconductor film. It aims to improve charge-carrier transport between cQDs through the replacement of long, insulating molecules by short, conductive ones that can still passivate the surface. This process will determine two main measurable effects: the volume contraction due to a reduced inter-particle distance, and the disappearance of vibrational features associated to the removed molecules. AFM-IR is ideally suited for this study as it allows the measurement of both effects simultaneously at the nanometre scale. This technique is able to overcome resolution limitations associated with conventional IR spectroscopy, as the wavelength is on the same size scale as the analysed microstructures (Figure 1(a)). The metallic AFM tip first enhances the incoming IR laser pulses, resulting in a local excitation of vibrational modes in the OA ligands (Figure 1(b)). The consequent pulsed thermal expansion of the sample is proportional to the absorption coefficient, and can be detected mechanically by the same AFM tip.

Carboxyl groups in OA are identified through the stretching mode of the carbon-oxygen double bond, between 1250 and 1650 cm$^{-1}$ when the molecule is bound to the PbS cQDs surfaces.[21,22,25] The AFM-IR mapping combines topographic and chemical analyses, identifying the IR-active regions on the sample (Figure 1(c)). The AFM-IR spectra are then obtained from the mapped structures (Figure 1(d)). The results match the corresponding FTIR spectra, with no significant difference between the EHD-printed structures when compared to the spin-coated PbS cQDs. Below, quantitative analysis of ligand exchange was performed by studying variations in the most intense signal (1410-1420 cm$^{-1}$). The results of each measurement were normalised against a spin-coated reference sample to account for variations related to the experimental setup only (see Methods for more details). The vibrational signal is still detectable upon the application of an EDT ligand exchange treatment, even if drastically reduced (Figure 1(e)). It is thus possible to compare the efficiency of different treatments based on this technique.

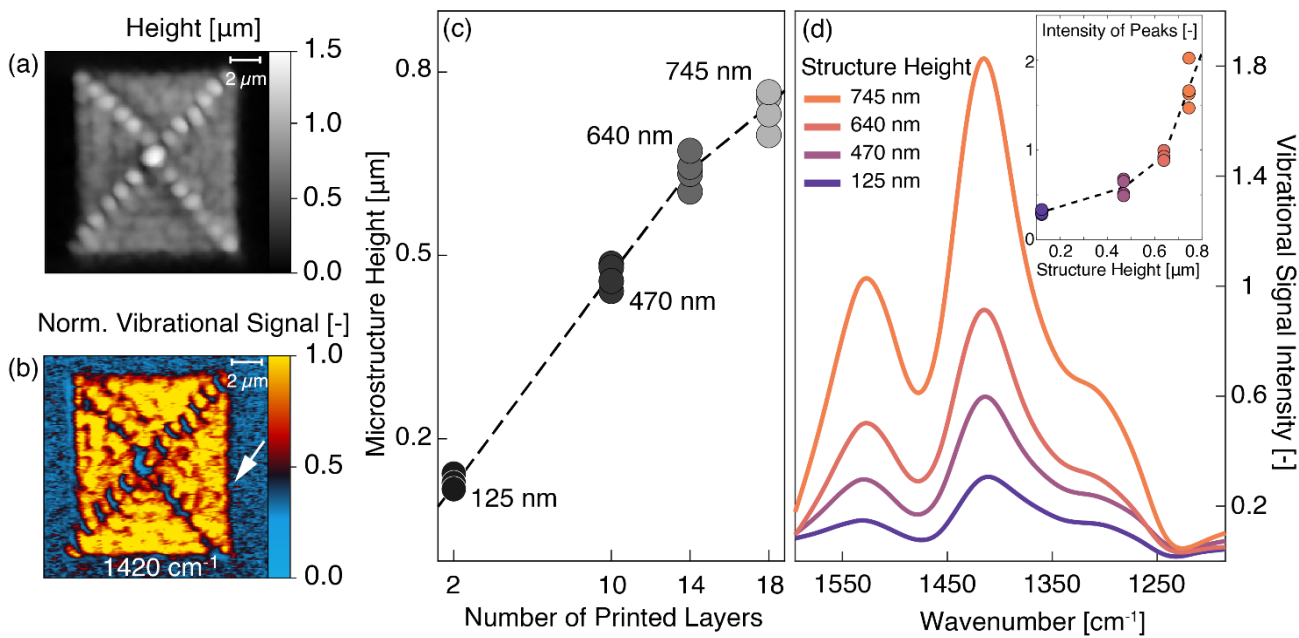

**Figure 2:** Topographic and vibrational information from EHD-printed PbS cQDs structures before ligand exchange. (a) AFM-IR topography of a representative printed structure. A thicker, cross-shaped feature can be noticed, unintentionally introduced due to the chosen

printing procedure. (b) Normalised AFM-IR vibrational signal map, demonstrating the presence of resonating molecules only in the microstructure. The cross-shaped feature shades some regions from the excitation pulses, determining variations in signal intensity, as noticeable in Figure 1(c). The white arrow shows the direction of the laser pulses used for excitation, consistent with the shading. (c) Measured heights of the printed microstructures analysed in this work. Written values represent averages. (d) Vibrational spectra collected from the same printed structures and normalised as described in the methods. The inset shows the proportionality between the main peak intensities and the height of structures. For both (c) and inset (d), the average values are linked by the black dashed line as a guide to the eye.

The AFM-IR scans show how the chosen printing path results in an overall uniform film, while introducing a cross-like feature where excess cQDs are deposited (Figure 2(a)-(b)). The height values considered for this work neglect these features, and only refer to the underlying plateaux. These features have an impact on the vibrational mapping, as they shade neighbouring regions from the IR excitation, determining variations in the intensity of the vibrational signal related to the different excitation intensities, as reported in literature[26] (Figures 1(c) and 2(b)).

The EHD-printed structures with 2, 10, 14, and 18 printing loops of cQDs correspond to well-resolved average heights of 125±20, 470±60, 640±75, and 745±90 nm respectively (Figure 2(c)). The roughness of the microstructure surface increases with the number of printing loops; every layer interacts with the previous one due to the transitory presence of the solvent, leading to larger roughness and dispersion of height values.

As expected, the height of the printed features is directly proportional to the intensity of the vibrational signal, as more functional groups are excited and interact with the AFM tip (Figure 2(d)). The penetration depth of the IR light pulses used for the excitation of ligands in the printed structures is larger than 745 nm, as no saturation of the vibrational signal intensity can be discerned with increasing structure height (Figure 2(d) inset). The obtained dependence can be used as calibration curve to measure the amount of ligands left after the different ligand exchange procedures.

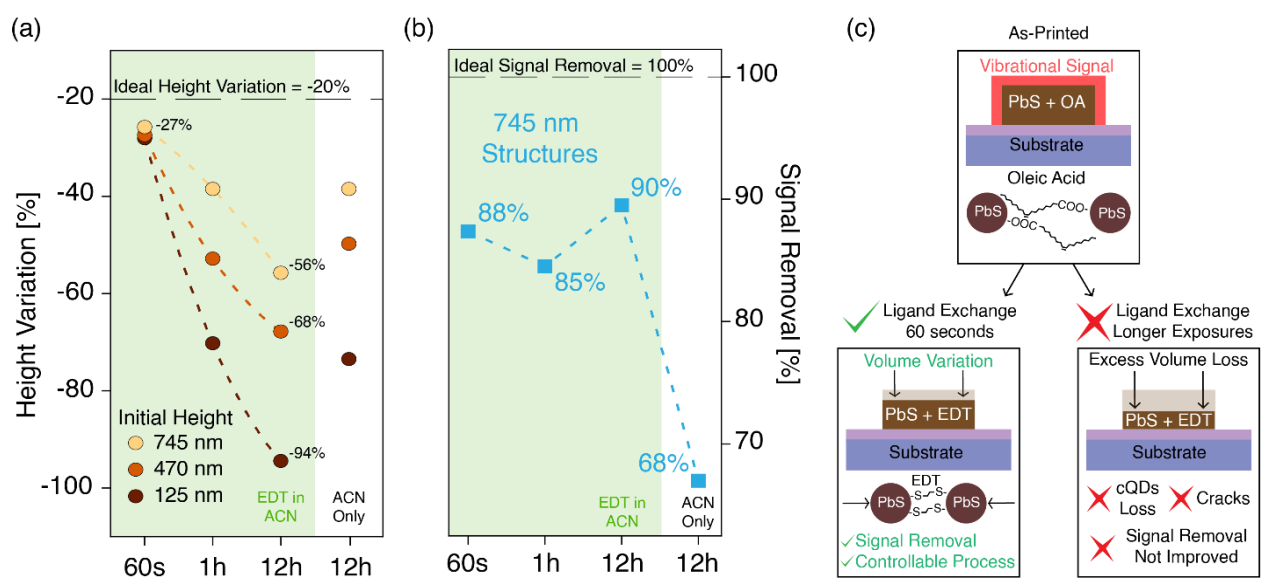

**Figure 3:** Volume loss and signal removal upon different ligand solution exposure times. The reported variations are calculated based on average values. (a) Height loss in structures with different initial height. The ideal level of -20% volume loss is represented as dashed horizontal line. The shortest treatment introduces uniform volume losses of -27% in different initial volumes, suggesting a controllable ligand exchange process. Longer treatments introduce excessive losses, with a dependence on the surface-to-volume ratio of structures:

the larger the ratio, the larger the losses. The reduced losses in absence of EDT molecules (ACN-only treatment) are related to the milder interaction with the pure solvent, but are still dependent on the surface-to-volume ratio, and consistent with the loss of active material. (b) Oleic acid vibrational signal removal upon ligand exchange. The ideal level of 100% signal removal is represented as dashed horizontal line. More than 85% of the OA signal is removed by any of the EDT treatment, independently of the exposure time. The ACN-only treatment does not remove the OA-vibrational signal as effectively, demonstrating the importance of EDT concentration in the process. (c) The illustration represents how the ideal treatment results only in substitution of the molecules, associated with the disappearance of vibrational signal from carboxyl groups and the reduction of interparticle distance, causing volume loss in structures. Treatments with excessive exposure time result in degradation of the active structures, with no improvement of the OA removal.

The overall quality of the ligand exchange process is studied here, considering both volume losses and vibrational signal variations. The structures are exposed to the same ligand solution for increasing times (60 seconds, 1 hour, and overnight (12 hours)), to evaluate if a short interaction time is a limit for a complete exchange, and if long exposures improve results or rather introduce damage. Three structure heights are investigated (125, 470 and 745 nm) to assess the role of the initial volume on the interaction with the ligand solution. The exposure to ACN overnight is included as control experiment to decouple the effect of ligands and solvent.

Volume variations in the microstructures are expected, following the reduction of PbS cQDs inter-particle distance under ligand exchange from OA to EDT. These variations can be modelled through the contracting-random-loose-packing model, developed for printable inks of metallic nanoparticles.[27] Different inter-particle distances have been reported for PbS, depending on whether ligand shells partially merge[28] (from 2.6 nm with OA to 1.2 nm with EDT) or remain fully isolated[29] (from 4 to 1.6 nm).[30] Considering the case of partially-merged ligand shells, and PbS cQDs with a core diameter of 6 nm, a complete ligand exchange would correspond to a 20% volume loss.

In this work, the volume loss is directly translated in height loss, considering for simplicity the in-plane shrinkage as negligible. Topography scans are used to measure the height variations and to check for structural damage. The shortest ligand treatment of 60 s is the most controllable and homogeneous one, resulting in an equal height loss of 27% across all printed microstructures, independently of their initial volumes (Figure 3(a)). This is consistent with the replacement of ligands, plus a minor degree of cQD loss, pushing the volume loss only slightly beyond the expected 20%.

Excessive volume losses are measured when the cQD structures are exposed to the ligand solution for longer periods of time. The interaction with the ligand solution in this case is correlated with the surface-to-volume ratio of the microstructures. When left overnight in the EDT solution, the 125 nm structures are almost entirely removed (-94%), while 470 and 745 nm structures lose 68% and 56% of their volumes, respectively. These volume losses and their dependence on the surface-to-volume ratio are consistent with the removal of active material from the microstructures, detrimental for device performance. These same conditions also result in the formation of cracks (Supporting Information Figure S3), possibly creating short circuits if such printed structures are then integrated in vertical devices. Neglecting the in-plane shrinkage with this extent of structural damage results in a conservative estimate of the total volume loss. Such damages are instead not visible in any microstructure when the 60 s ligand treatment is applied.

In the case of overnight exposure to ACN the interaction is milder, but results are still consistent with the loss of active material from the microstructures: percentages exceed the ones obtained from modelled ligand removal processes, and a dependence on surface-to-volume ratio is still present. Although some reports showed that ACN does not interact with the ligands[31], this solvent has been chosen previously to optimise SSLE processes because of its interactions with the cQD films. In particular, cQDs can rearrange themselves in the layer during post-deposition ACN treatments, leading to a self-curing process of cracks and defects.[32] The longer exposure time applied in this work instead drives solvent degradation of the photoactive cQD microstructures, with a percentage of the active material being removed.

The information on volume loss must be coupled with the measurement of the OA vibrational signal removal to determine the efficiency of the ligand exchange process. The analysis is focused on the structures with an initial height of 745 nm, being the most relevant case for photodetector applications requiring efficient light absorption, thus thicker active layers. Within 60 s, 88% of the signal is removed from a 745 nm structure, and the signal loss is larger than 85% for any treatment duration (Figure 3(b)). When considering the control sample exposed to ACN only, the removal of the vibrational signal is less effective, reaching only 68%. This behaviour, showing little oscillation around the same value in a large range of exposure times, and a clearer variation upon the change of ligand concentration, is consistent with previous reports. A sigmoidal dependence of properties on the concentration of the ligand solution was demonstrated earlier[33], suggesting that the timescale analysed in this work, together with the constant concentration of ligands, leads to consistent results. The analysis of shorter timescales, together with ligand concentration variations, could help understanding these processes further.

In conclusion, AFM-IR allowed us to quantitatively assess the ligand exchange efficiency in the challenging case of EHD-printed microstructures of PbS cQDs, beyond the capabilities of conventional FTIR measurements. For this, 10×10 µm$^2$ structures with varying volumes were analysed by simultaneously investigating their topography and vibrational signatures. Considering structure with heights approaching the micrometre scale (745±90 nm), the preferred treatment is the shortest. Only 60 s are needed to remove approximately 90% of the vibrational signal, without introducing excessive active material losses or structural damage. Longer exposures to the ligand solution incur deviations from the modelled volume loss, corresponding to cracking and removal of the photo-active cQDs, without improving the substitution of molecules. Excessive volume loss and cracking can also be caused by overnight exposure to ACN, suggesting that it damages the active layer without removing oleic acid. Through the combination of microscale additive manufacturing and nanoscale quantitative analysis, this work provides important knowledge towards the development of solution-processed devices based on conductive colloidal semiconductors.

**Methods**

*PbS cQD synthesis*

PbS QDs were synthesized with slight adaptions to the method described by Hines et al.[34] as described previously.[19] Spectra of the PbS cQDs dispersed in tetrachloroethylene were acquired by UV-Vis spectroscopy (Jasco V-670).

*Spin coating of reference sample*

The as-synthesized PbS cQDs were redispersed in octane (240 mg/ml) and filtered with a 0.1 μm PTFE syringe filter. The drop was placed on the substrate while spinning it at 1500 rpm for 45 s.

*EHD printing*

The as-synthesized PbS cQDs were redispersed in n-tetradecane at a concentration of 40 mg/ml and filtered with a 0.1 μm PTFE syringe filter. They were then deposited using electro-hydrodynamic printing (EHD) with a commercial system (Scrona NanoDrip™ R&D Print System and a Gen11p51 print head).

The Scrona print head is a MEMS print head manufactured from silicon wafers. In comparison to single-nozzle capillary-like print heads used in many other studies voltage is not only applied between the substrate and the nozzle but there is an additional electrode embedded in the print head itself that is situated between at a small distance from the nozzle, facing the substrate chuck. The voltage Vextraction applied to this electrode ("nozzle Voltage") will cause droplet ejection if the difference to the voltage Vink applied to the ink ("PH Voltage") is sufficiently large. As the droplets are ejection and pass across the extraction electrode they will be guided straight downwards given the uniform electric field that is formed between the approximately 1mm wide extraction electrode and the chuck which is driven at Vchuck. In this way, even at the used printing distance of 500 μm to the substrate accurate placement is secured.

The SiOx sample (525 μm thick Si wafer with 280 nm Oxide) was glued and grounded using silver paste onto an ITO glass substrate, which was grounded to the printing system. The starting off-state voltage levels which are applied to precondition the ink meniscus are the following: 500 V at chuck, 150 V at nozzle, 20 V at print head (PH), 3 ms pulse duration, with nozzle pulse off duration set to 0 ms. In this way, the effective applied voltages are switched each 3 ms from 500 V chuck, 300 V nozzle, -250 V PH to -500 V chuck, -300 V nozzle, 250 V PH. The distance between print head and ITO glass next to sample was set to 500 μm. The system employs two complementary objectives: a bottom optic positioned beneath the substrate, enabling in-situ process monitoring exclusively with transparent substrates, and a top optic for post-process print examination. As our substrate was non-transparent, we initially calibrated the printing process on transparent ITO material before transitioning to the non-transparent substrate. Printing was initialized by increasing nozzle and PH voltage (to ~300 V nozzle, ~250 V PH) on ITO glass until a spot size of ~2 μm was achieved (observed with bottom microscope). When a stable printing pattern could be achieved, the print head was moved to the substrate with same relative distance to its surface (checking the orientation with top microscope), and printing started with +100 V higher chuck voltage (the oxide layer allows higher chuck voltages applied, resulting in better printing quality).

*Ligand Exchange Procedure*

The native ligands were exchanged by a solid-state ligand exchange treatment with 2% vol. ethane-1,2-dithiol (EDT) in acetonitrile. For the 60 s treatment, the sample was placed on a spinner and fully covered by a drop of EDT in acetonitrile (~ 150 µL). After 60 s, the sample was spin-dried at 2500 rpm for 45 s. The sample was rinsed with 5 drops of pure acetonitrile while spinning at 2500 rpm for 45 s. For the longer treatments, the samples were soaked in EDT in acetonitrile for 1 h or overnight and pure acetonitrile, respectively. The EDT-treated samples were then dipped in pure acetonitrile and blow-dried with $N_2$. Subsequently, all the samples were placed on a spinner and rinsed with five drops of pure acetonitrile while spinning at 2500 rpm for 45 s.

*AFM Structure Height Measurements*

An Anasys nanoIR2 platform was used for both topographic and vibrational analysis of the printed structures. Each of the printed structures was scanned in tapping mode, as the active material is relatively soft and may be damaged when other scanning modes are used. The measured heights were obtained by considering the histogram of height values for each structure, with the first two main peaks representing the substrate and the main plateau of structures. The difference between these two peaks is reported here as the structure's height. In this way, the cross-like feature is neglected.

*AFM-IR analysis*

Vibrational signals were collected using an Anasys nanoIR2 in tapping mode, with the main parameters used to obtain the desired chemical contrast being IR tuneable laser power (5 %) and drive strength (20 %). Tapping strength and laser power are balanced to avoid either thermal or mechanical damage to imaged structures. Centre frequency and frequency range were in the expected range for tapping-compatible AFM-IR probes (PR-EX-TnIR-A10).

To perform a quantitative study of the vibrational data an analysis procedure was purposefully developed in this work, based on the comparison of every structure's spectrum with the spectra obtained from a reference spin-coated sample with a cQD layer thickness of 260 nm. The complete procedure of data collection and analysis is summarised in Figure S2. The reference sample, with a uniform surface of cQDs, provides signals with minimal variations across different measurement sessions due to the uniform distribution of OA capped cQDs, avoiding large variations between tip landing points, and to the low roughness, avoiding shading effects. Spectra are collected from the reference before and after every measurement on printed structures. The reference signal will then highlight variations resulting from unstable laser power, from progressive contamination of the AFM tip, or from misalignments developed during measurements. Variations in the reference signal were not constant, but could reach values of 20% between following measurements.

Every measurement, including the ones from the reference sample, is repeated several times, with the minimum being four. As both the vibrational and topographic nanoscale data collection happens through the AFM tip, all measurements are extremely sensitive to local properties, which may not be representative of the whole structure. Multiple measurements on a micrometric region considered to undergo homogeneous experimental conditions are thus collected to determine a structure's average state, in order to account for the large roughness of printed structures, which may determine variations in vibrational signal due to different amounts of OA molecules or to shading of the excitation source. As visible in Figure 1(d), the AFM-IR spectra are represented by smoothing splines (b-spline). The peak values, averaged between 1375 and 1450 $cm^{-1}$ used for ligand concentration in the cQD structures are extracted after the smoothing to account for the relatively large noise level in measured spectra. The points extracted from every measurement are then averaged, and the standard deviation is reported.

Acknowledgement

We are grateful to Dr Lily Poulikakos of Empa for the use of the AFM-IR, and to Patrick Galliker, Julian Schneider, and Anni Wang from Scrona for their support and assistance in identifying optimal printing parameters for our research. The authors acknowledge financial support from the Swiss National Science Foundation (SNSF, project no. 200021 182790).

# Supporting Information

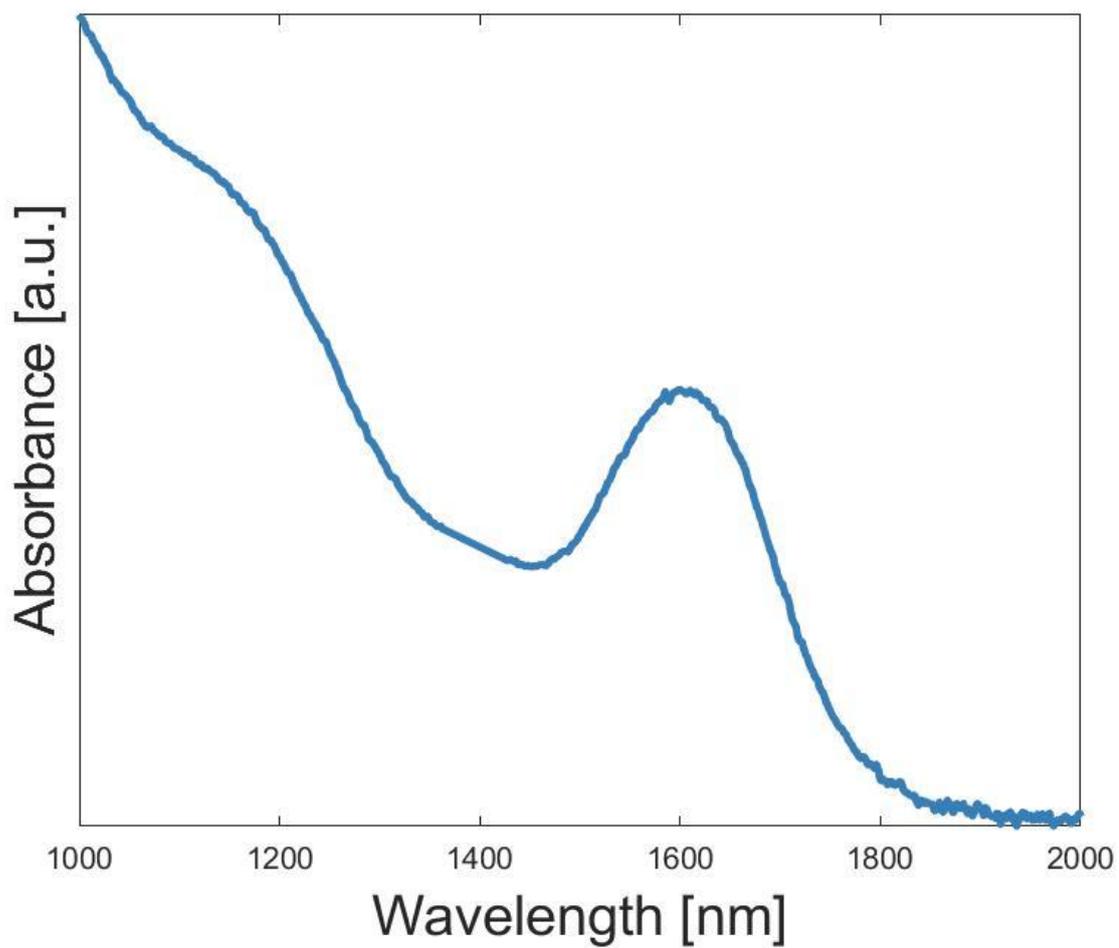

**Figure S1:** Absorbance spectrum of the PbS cQD solution used in this work, showing their well-defined excitonic feature around 1600 nm.

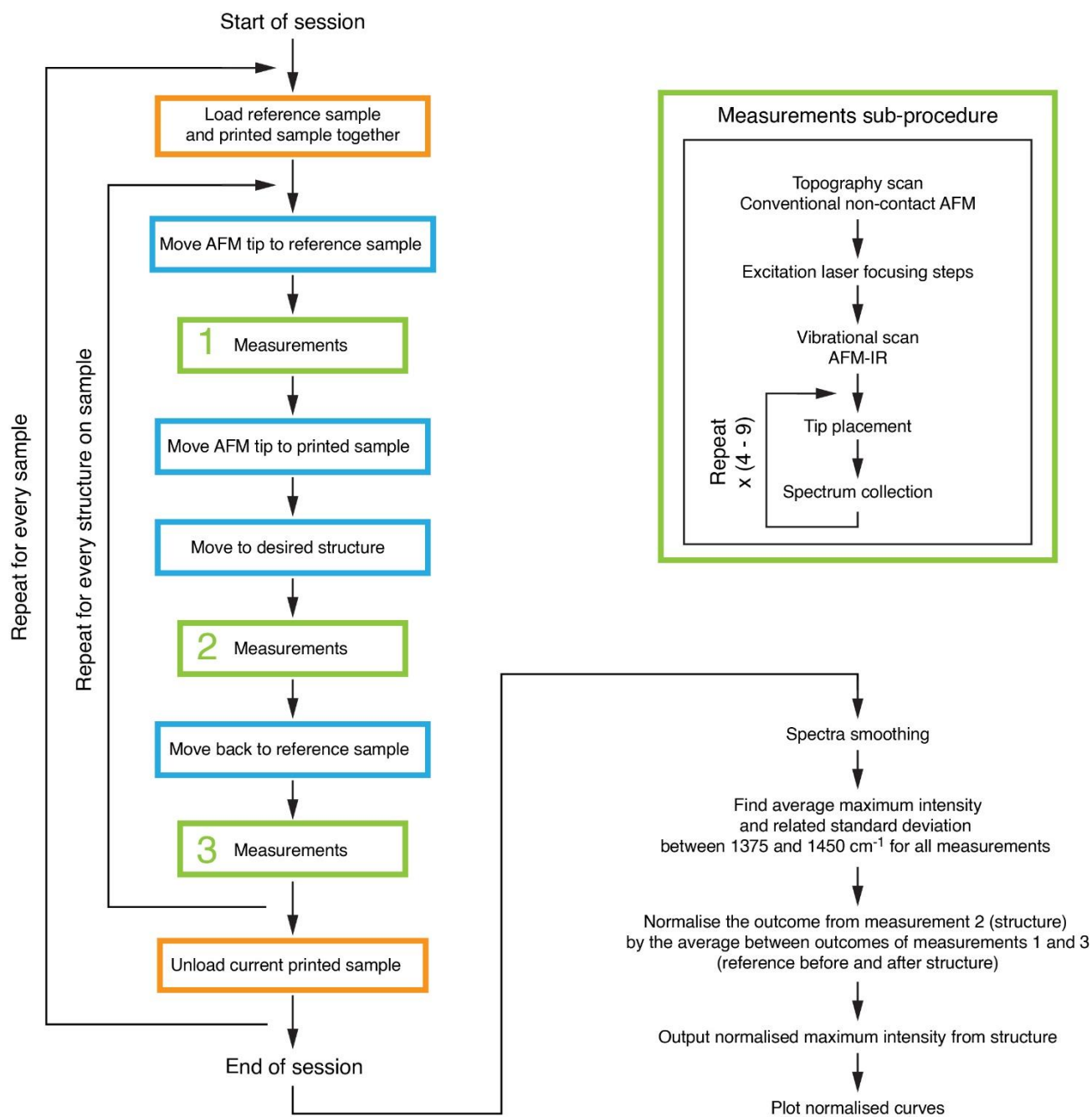

**Scheme S2:** Complete procedure for AFM-IR data collection and data treatment. Data collection steps are colour coded as follows: load and unload (orange), movements of AFM tip (blue), measurements (green). Top right inset shows the sub-procedure of each measurement step.

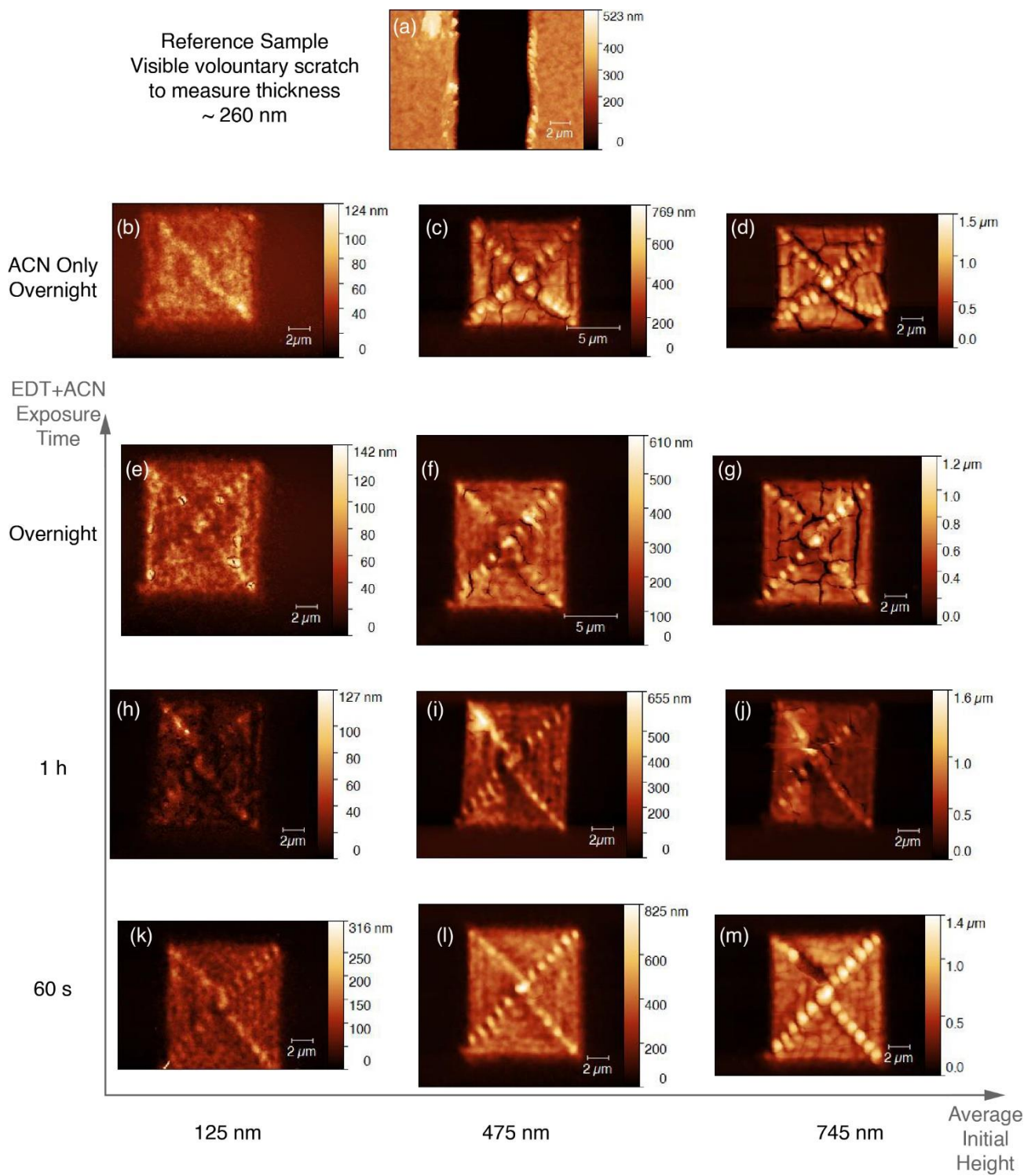

**Figure S3:** AFM topography scans of every analysed structure after ligand exchange, including reference sample. Every row corresponds to a sample, and in printed sample increasing printed structure heights are found from left to right. Refer to Figure 3(a) for the height variation introduced by the ligand exchange. (a) Spin-coated reference sample with scratch voluntarily introduced to expose the substrate and measure the thickness of the ligand-exchanged cQD film. (b)-(d) Sample exposed to ACN only. (e)-(g) Sample exposed to EDT in ACN overnight. (h)-(j) Sample exposed to EDT in ACN for 1 hour. (k)-(m) Sample exposed to EDT in ACN for 60 second. The large defect in (m) on the top-left arm of the cross-like structure has been introduced during measurements, and is caused by mistakes in handling of samples.

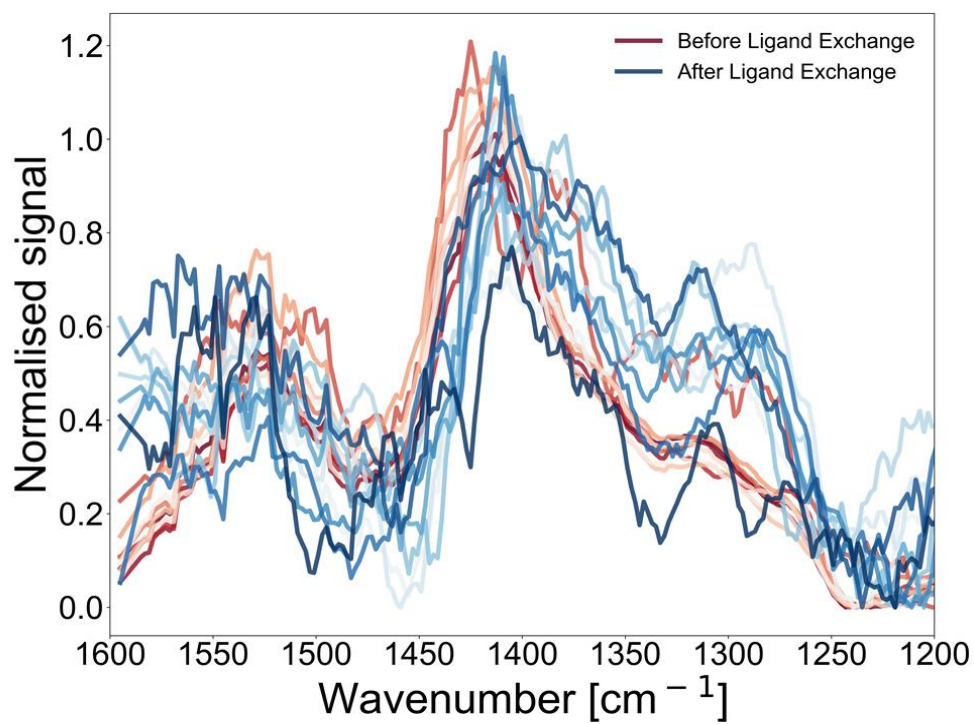

**Figure S4:** Comparison between normalised AFM-IR spectra before (red) and after (blue) ligand exchange. The measured structure has average height of 745 nm, on the sample exposed to EDT in ACN for 60 seconds. Spectra have the same profile, but the signal to noise ratio is lower after ligand exchange as ~90% of the signal is removed. The same spectra normalised to the reference sample only are visible in Figure 1(e).

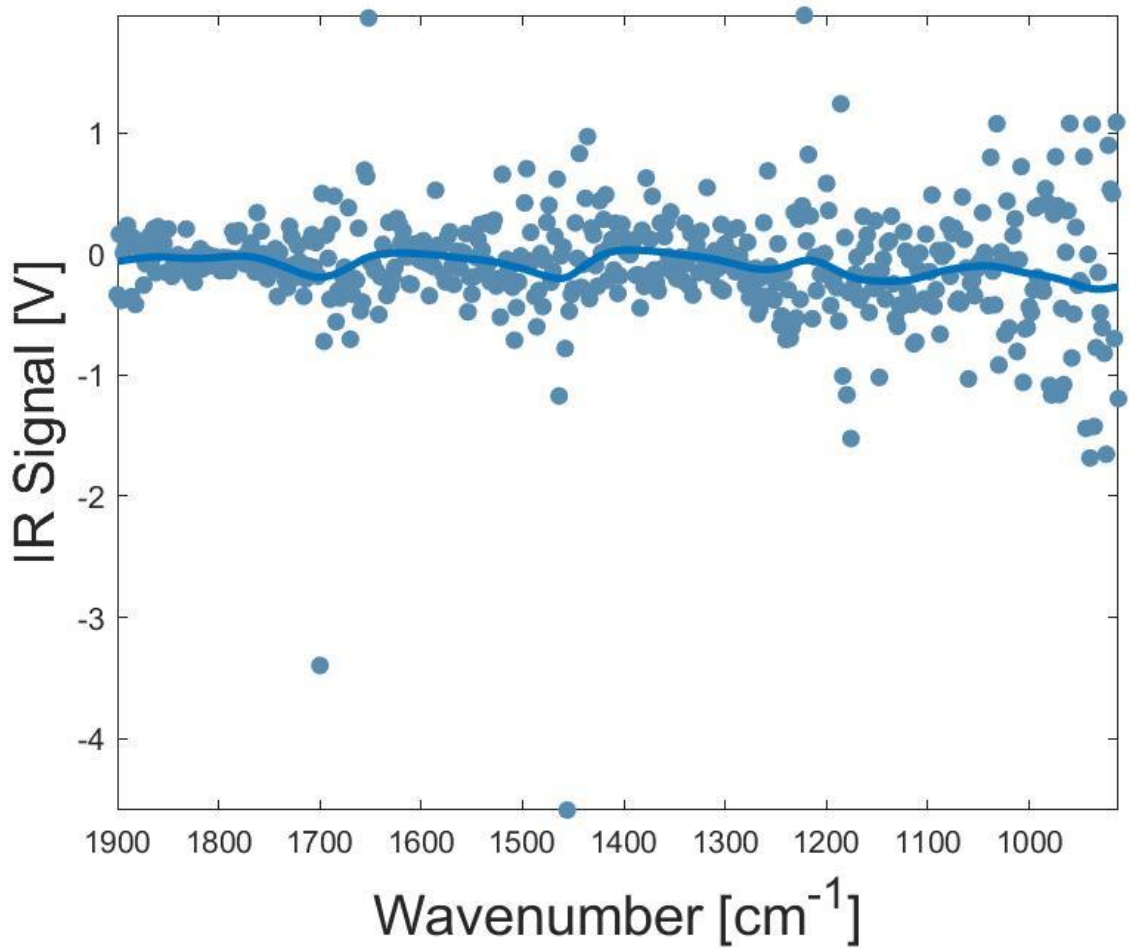

**Figure S5:** AFM-IR vibrational spectra measured from the substrate of PbS cQDs structures (silicon dioxide). No signal can be highlighted, as values of intensity remain close to zero. The oscillations are due to the excitation source, as four different lasers are needed to cover the full wavenumber range, and the transitions between them are visible in the detected signal. Intervals are the following (values expressed in $cm^{-1}$): 902-1180, 1180.1-1449, 1449.1-1700, 1700.1-1958.

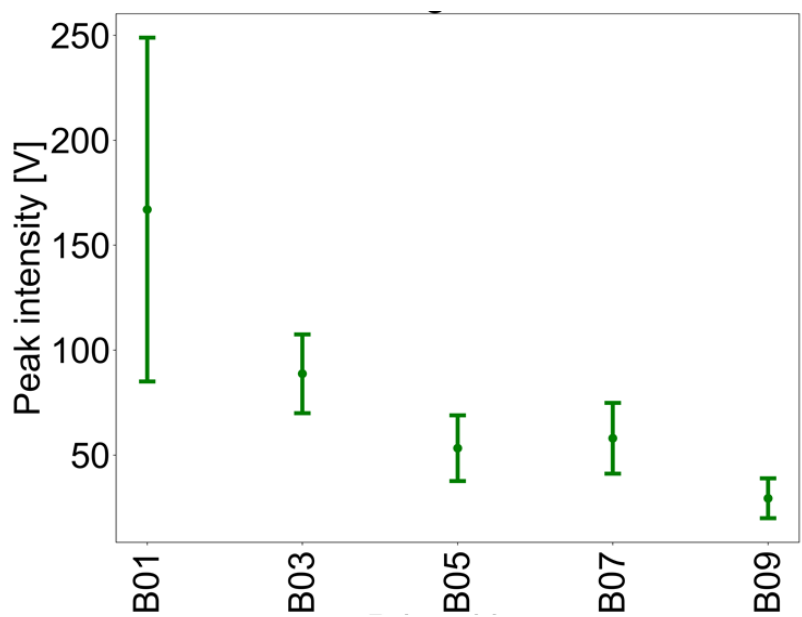

**Figure S6:** The average vibrational signal intensities (dots) and assigned error (bars) from the as-printed structures. The structure height decreases from left to right. The values are reported in volts as provided by AFM-IR setup (before normalisation with the reference sample).

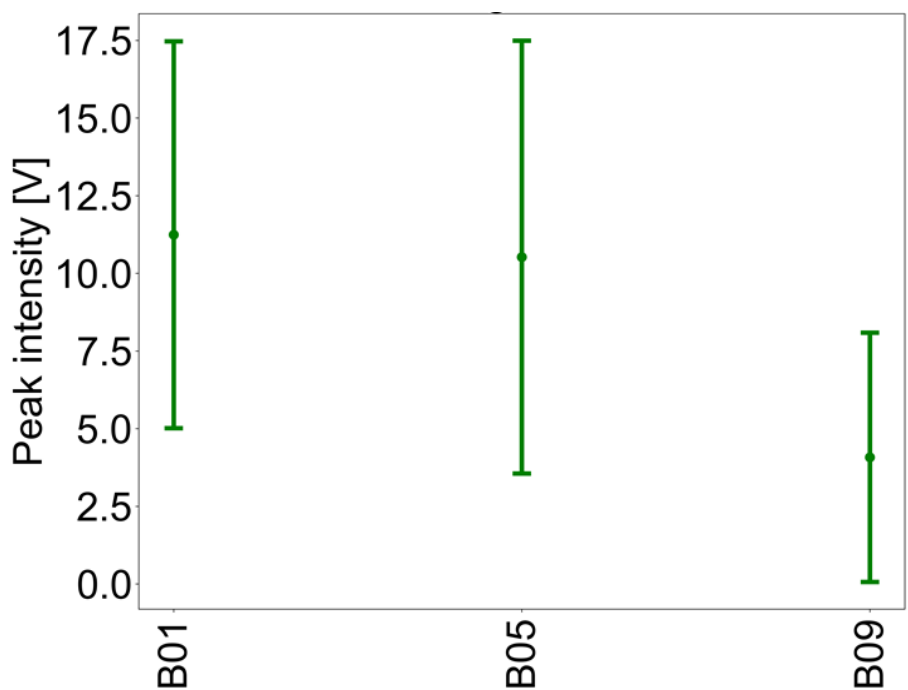

**Figure S7:** The average intensities and assigned error from 3 structures of Figure S6 exposed to EDT in ACN for 60 seconds. The structure height decreases from left to right. The values are reported in volts as provided by AFM-IR setup (before normalisation with the reference sample).

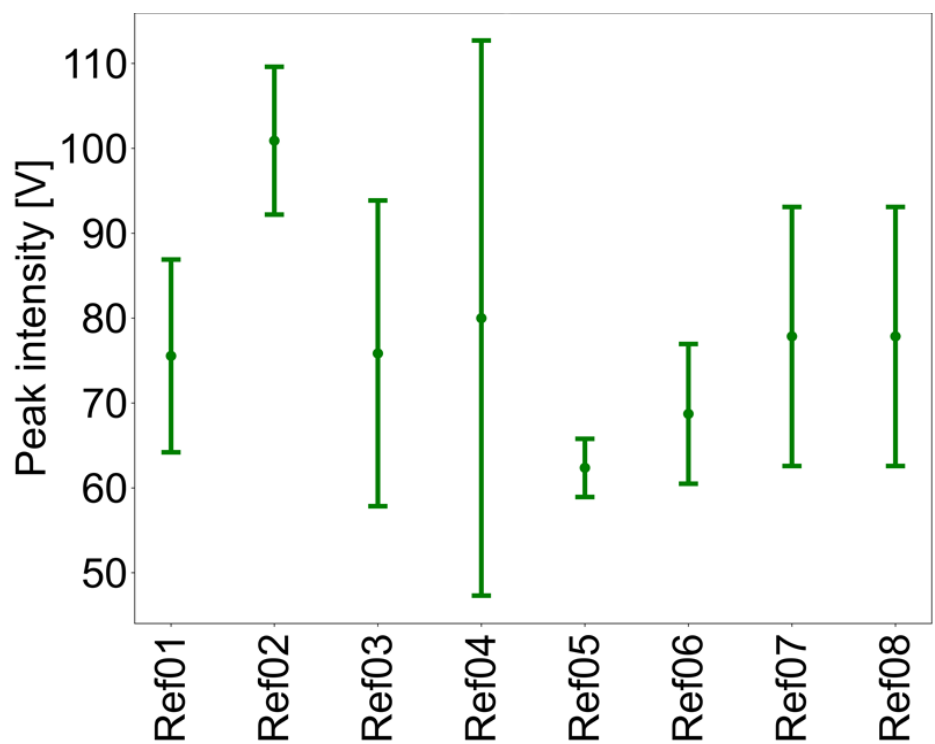

**Figure S8:** The vibrational signal from the reference sample used for normalisation along the measurement session of as-printed structures. The values are reported in volts as provided by AFM-IR setup.

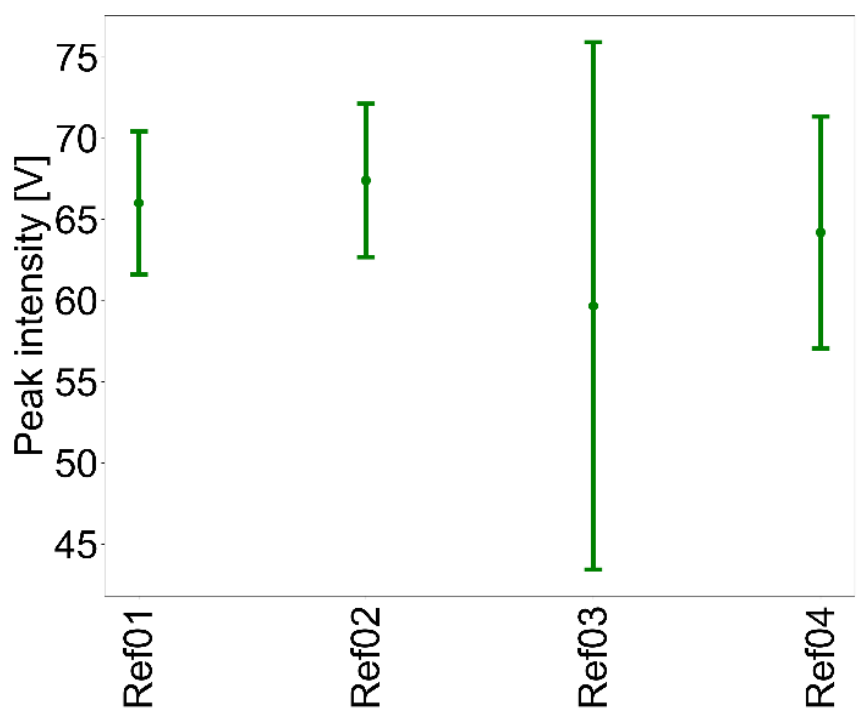

**Figure S9:** The vibrational signal from the reference sample used for normalisation along the measurement session of ligand-exchanged structures. The values are reported in volts as provided by AFM-IR setup.

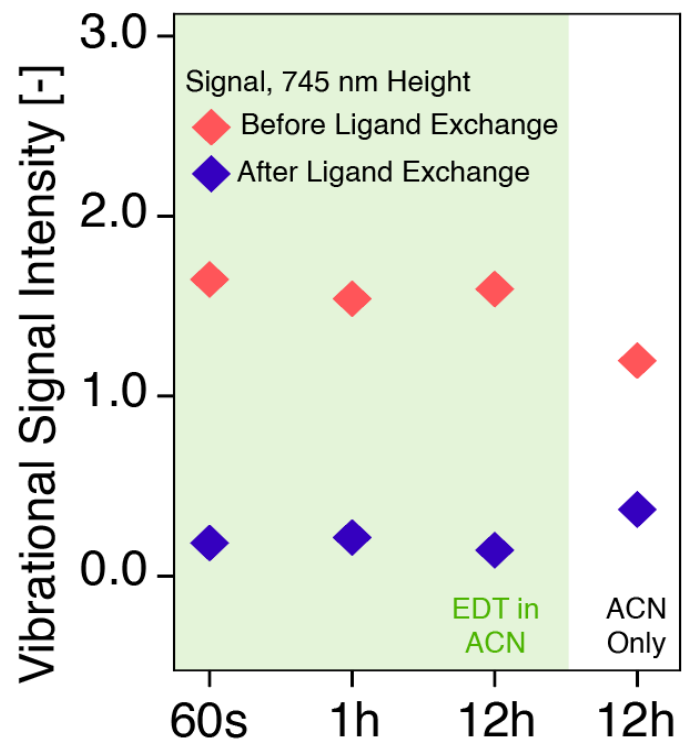

**Figure S10:** Variation of vibrational signal intensity for 745-nm-high microstructures upon the application of different ligand exchange procedures. Average values of intensities are reported.